\documentclass[reprint,aps,prb]{revtex4-2}
\usepackage{color,graphicx,calc,url}
\usepackage{dcolumn}
\usepackage{bm}
\usepackage{amssymb}
\usepackage{amsmath}
\usepackage{wasysym}
\usepackage{mathrsfs} 
\usepackage{gensymb}
\usepackage{amssymb}
\usepackage{wrapfig}
\usepackage{floatrow}
\usepackage[ppl]{mathcomp}
\usepackage{hyperref}
\usepackage{color}
\definecolor{blue}{rgb}{0,0,1}
\definecolor{red}{rgb}{1,0,0}
\definecolor{darkred}{rgb}{0.5,0,0}
\begin{document}
\title{Role of NiO in the nonlocal spin transport through thin NiO films on Y$_3$Fe$_5$O$_{12}$}
	

\author{Geert R. Hoogeboom$^1$, Geert-Jan N. Sint Nicolaas$^1$, Andreas Alexander$^2$, Olga Kuschel$^2$, Joachim Wollschl\"ager$^2$, Inga Ennen$^3$, Bart J. van Wees$^1$\ and Timo Kuschel$^{3}$ \email{Electronic mail: g.r.hoogeboom@gmail.com}}
\affiliation{$^1\,$Physics of Nanodevices, Zernike Institute for Advanced Materials, University of Groningen, Nijenborgh 4, 9747 AG Groningen, The Netherlands,\\
$^2\,$Department of Physics and Center of Physics and Chemistry of New Materials, Osnabrück University, Barbarastraße 7, 49076 Osnabrück, Germany\\
$^3\,$Center for Spinelectronic Materials and Devices, Department of Physics, Bielefeld University, Universitätsstraße 25, 33615 Bielefeld, Germany}\

\date{\today}
\begin{abstract}
In spin transport experiments with spin currents propagating through antiferromagnetic (AFM) material, the antiferromagnet is treated as a mainly passive spin conductor not generating nor adding any spin current to the system. The spin current transmissivity of the AFM NiO is affected by magnetic fluctuations, peaking at the N\'eel temperature and decreasing by lowering the temperature. 
In order to study the role of the AFM in local and nonlocal spin transport experiments, we send spin currents through NiO of various thickness placed on Y$_3$Fe$_5$O$_{12}$. The spin currents are injected either electrically or by thermal gradients and measured 
at a wide range of temperatures and magnetic field strengths. The transmissive role is reflected in the sign change of the local electrically injected signals and the decrease in signal strength of all other signals 
by lowering the temperature. 
The thermally generated signals, however, show an additional upturn below 100$\,$K which are unaffected by an increased NiO thickness. A change in the thermal conductivity could affect these signals. 
The temperature and magnetic field dependence is similar as for bulk NiO, indicating that NiO itself contributes to thermally induced spin currents. 

\end{abstract}

\maketitle
\section{Introduction}


The scope for using antiferromagnets (AFMs) as base for spintronic devices has been unveiled by the possibility to inject spin angular momentum \cite{Wang2014,Lin2016,Prakash2016,Wu2016} which can be carried over long distances \cite{Lebrun2018a,Yuan2018,Xing2019,Hoogeboom2020,Wimmer2020,Han2020} and by the read-out of the magnetic order \cite{Hoogeboom2017,Fischer2018,Baldrati2018}. AFM dynamics is fast compared to that of ferromagnets (FMs) due to high eigenfrequencies \cite{Johnson1959,Hutchings1972}. Further, AFMs have vanishing magnetization which increases its robustness against magnetic perturbations and reduces the cross talk between AFMs \cite{Loth2012}. The manipulation of the magnetic moments thus requires relatively high magnetic fields as compared to FMs. However, the needed magnetic field strength to control the magnetic alignment in the AFM can be reduced by combining a thin AFM layer with FM layers due to the exchange coupling across the interface. 



Spin current has been sent efficiently normal through the AFM$|$FM bilayers NiO$|$YIG (yttrium iron garnet, Y$_3$Fe$_5$O$_{12}$) in a local geometry for which the relaxation length in NiO is a few nm \cite{Wang2014,Hahn2014a,Lin2016,Qiu2016,Li2016,Prakash2016,Lin2017,Hou2017,Hung2017a,Luan2018a}. 
The spin currents 
can be generated with electrical means via the spin Hall effect (SHE) in Pt causing electron spins to accumulate at the Pt$|$NiO interface, or by a heat gradient throughout the magnet via the spin Seebeck effect (SSE). The spin current through the interface is established via a finite spin mixing conductance and spins flowing into the Pt are converted into a charge current by the inverse spin Hall effect (ISHE). 

Spin Hall magnetoresistance (SMR) is the local combination of the SHE and the ISHE and is sensitive to the applied spin transfer torque on the magnet. This depends on the direction of the magnetic moments under influence of a magnetic field. 
Lowering the temperature of these devices results in a sign change in the SMR showing that the spin current interacts with magnetic moments that are 90$\,\degree$ angular shifted as compared to room temperature.
Although the magnetic moments of YIG align parallel to the magnetic field, those of NiO tend to align perpendicular to the magnetic field to lower the Zeeman energy and due to exchange bias. Therefore, NiO can be the source of this angular shifted, negative SMR \cite{Hou2017,Luan2018a} which angular dependence resembles bulk NiO \cite{Hoogeboom2017,Fischer2018}.

In addition to spin transfer torque which is damped out over short distances, the spin accumulation in Pt also causes magnetic excitation in the magnetic bilayer; magnons.
These quasiparticles carry the spin angular momentum over long distances. Furthermore, the creation of magnons with Joule heating from the charge current in Pt causes the magnons to flow from the hot towards the cold part. This results in a negative magnon chemical potential $\mu_m$ near the injector and a positive $\mu_m$ at some distance from it 
\cite{Shan2016}.

\begin{figure*}[t]
\includegraphics[width=18.0cm]{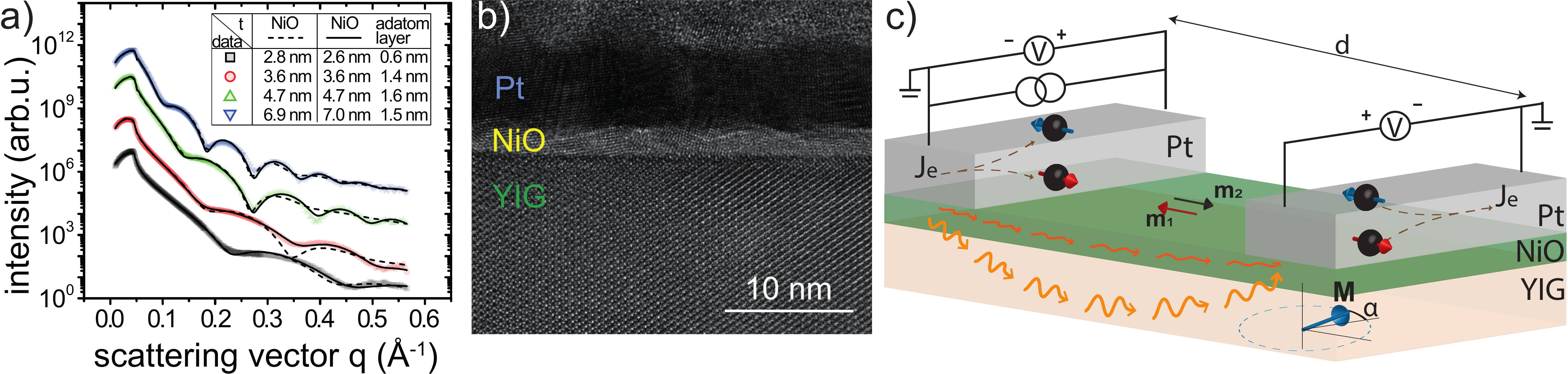}
\caption{a) X-ray reflectivity data and fits of exemplary samples with various NiO thicknesses $t$ prior to surface treatment and Pt device deposition. The data is vertically shifted for clarity. The black lines are fits. The dashed lines consider a single homogeneous NiO layer while the solid lines allow the existence of an additional adatom layer. The resulting fit parameters for the thickness of NiO and the adatom layer are included in the inset table for both fit methods which result in almost the same NiO thicknesses. All fits consider the YIG layer to be infinite and the GGG substrate to be negligible due to the relative thick YIG layer of 260$\,$nm. b) TEM image of the NiO(4.9$\,$nm$^*$) device. The cross section shows the Pt injector and NiO thin film being polycrystalline as well as the single crystalline YIG. c)  Illustration of the device structure including the electrical injection measurement scheme. The YIG magnetization $M$ is rotated in-plane by a magnetic field with angle $\alpha$. By the exchange bias with YIG, the NiO magnetic moments in both sublattices align perpendicular to the magnetic field. The SHE-induced spin accumulation in Pt causes transfer of spin angular momentum into the magnetic bilayer by the spin-flip mechanism. The injected spin current is carried by propagating magnons through the NiO$|$YIG bilayer and detected as a voltage via the ISHE in a second Pt strip at distance $d$.}
	{\label{fig:device}}
\end{figure*}

All these different sources and forms of spin current are influenced by the temperature. The positive SMR indicates that the NiO spin transmissivity is high at high temperatures. The interactions with magnetic moments in YIG dominate the effect on the spin accumulation over those in NiO. The SSE also shows this transmissivity effect as a peak at the N\'eel temperature
. These observations might 
%
%
be explained by magnetic fluctuations, giving the magnetic moments in NiO a component along the YIG magnetization and allowing the transport of spins along this direction through the AFM. 

In this article, we investigate the temperature dependence of different spin currents through the NiO$|$YIG bilayers in both local and nonlocal geometry. Most observations can be explained by a decrease in spin current transmissivity of the NiO layer below the N\'eel temperature .
Both local and nonlocal SSE signals show damping effects while lowering the temperature. The nonlocal electrically injected magnons pass the NiO layer twice and are therefore damped out stronger. Below 50$\,$K, however, all SSE signals show an upturn with decreasing temperatures. 
%
The temperature dependence of the thermal conductivity of the materials involved could change the thermal profile at low temperature. However, the low-temperature SSE signals are not affected by NiO thickness and therefore seem unrelated to the NiO transmissivity. In addition, the increase with magnetic field strength at low temperature is similar to that in bulk NiO, indicating that NiO itself contributes to the SSE.

\begin{figure*}[t]
	\includegraphics[width=17.0cm]{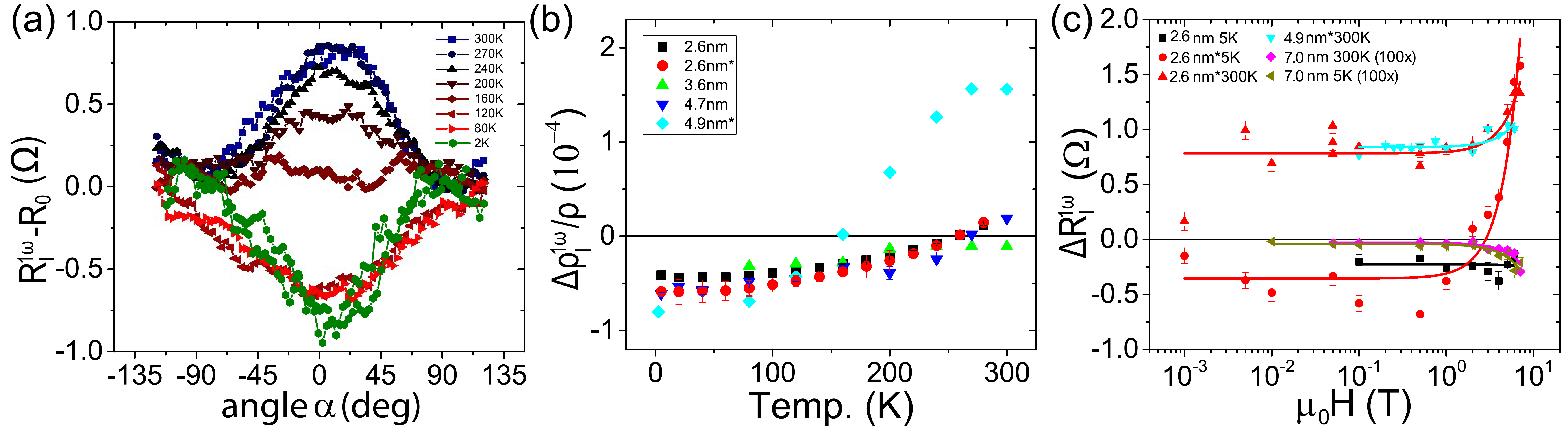}
	\caption{First harmonic local resistivity changes R$_l^{1\omega}$-R$_0$ obtained for the Pt$|$NiO(4.9$\,$nm$^*$)$|$YIG sample at 2$\,$T a) as a function of the rotation angle for various temperatures. R$_0$ is defined as the fitted background resistance and is zero at $\alpha=90\degree$.	Minor background signals due to temperature drift as a function of the time are subtracted. 
	The SMR modulation of the resistance shows an angular shift of $90\degree$ (negative SMR) by decreasing the temperature. b) The amplitude of the sinusoidal fit $\Delta \rho_{l}^{1\omega}/\rho$ is depicted as a function of the temperature for  the samples with various thicknesses of NiO showing the sign reversal at temperatures of 150$\,$K or above. c) $\Delta R_{l}^{1\omega}$ as a function of the magnetic field strength. The signals show quadratic magnetic field dependence as the corresponding lines of the fits show. For thick films, the quadratic increase becomes more negative, similar to films without YIG \cite{Fischer2018}. For thin films, the quadratic increase is positive, showing increased spin conductivity with larger magnetic fields.}
	{\label{fig:SMR}}
\end{figure*}

\section{Method}

The YIG films of 260$\,$nm thickness are obtained commercially and grown on a gadolinium gallium garnet (111) substrate. These were covered by NiO films of various thicknesses by reactive molecular beam epitaxy at a sample temperature of 250$\degree$C. Directly after deposition, the correct stoichiometry and chemical cleanliness of the films was checked in situ via x-ray photoelectron spectroscopy. To determine the exact thicknesses $t$ of the NiO films, the bilayer magnets were characterized by x-ray reflectivity (XRR) in a Philips X’Pert Pro diffractometer with a Cu K$_\alpha$ source. Exemplary curves are shown in Fig. \ref{fig:device}~a). The NiO optical parameters are fixed for the fits based on the Parratt algorithm \cite{Parratt1954} and taken from the Henke tables \cite{Henke1993}. The fit is improved when allowing an adatom layer with free optical parameters. However, this does not affect the NiO thickness significantly. 


To study the effect of magnetic order or a possible adatom layer on the interface spin transmissivity, some samples were etched by an Ar ion plasma etch at 200$\,$W for 10$\,$ seconds which is indicated as NiO($t$~nm$^*$). Sputter-deposited Pt strips were grown on top of these bilayer magnets by electron beam lithography using a 4$\%$ PMMA and an aquasave spincoat. 
Transmission electron microscopy (TEM) images have been taken in a JEOL
JEM 2200FS operated at 200\,kV using a Gatan OneView CMOS camera. The
TEM specimen has been cut out from the original structured sample by
focused ion beam preparation (FEI Helios NanoLab DualBeam). Figure \ref{fig:device}~b) shows a TEM image of the Pt(8$\,$nm)$|$NiO(4.9$\,$nm$^*$)$|$YIG(260$\,$nm) sample indicating a clean interface, a uniform thickness and a polycrystalline NiO structure. 

The Pt strips have dimensions of 20$\,\mu$m $\times$ 100$\,$nm $\times$ 8$\,$nm and the resistance between Pt strips was typically in the G$\ohm$ range. A high resistance is an indication of a relatively proper stoichiometry of the NiO \cite{Lu2002} and confirms any nonlocal signals are spin current related phenomena as opposite to amorphous YIG \cite{Perez2020}. Voltage spikes, however, occasionally resulted in a conductive film, making the devices unusable. Figure \ref{fig:device}~c) shows a schematic illustration of the resulting device structure including the electrical measurement setup. An alternating current of 100$\,\mu$A was sent through the injector strip and both the local and nonlocal voltage response is measured while applying a magnetic field in various in-plane directions~$\alpha$. 
The measurements are performed with a superconducting magnet system with a variable temperature inlet (VTI) and an ac lock-in method to distinguish electrical injected (first harmonic) and heat related (second harmonic) signals \cite{Vlietstra2014}. 

%






\section{Background}

Magnetic moments on neighboured (111) planes in NiO align antiparallel due to the relatively strong superexchange H$_e$ of 968$\,$T via the non-magnetic O$^{2-}$ \cite{Hutchings1972,Rezende2019}. Due to the applied magnetic field, the canting angle of the magnetic sublattices $\theta =  \arcsin(H/2H_e) \ll 1\degree$ \cite{Gurevich1996}, resulting in a small gain in the Zeeman energy. When the Zeeman energy is comparable to the magnetic anisotropy energy, the magnetic moments can align with respect to the magnetic field, which is easiest within the magnetic easy plane of one of the many possible domains. Even though SMR measurements in bulk NiO show a saturation at 6$\,$T \cite{Hoogeboom2017}, indicating that the majority of the magnetic moments are coherently controlled by the magnetic field, a NiO film of 120$\,$nm thickness shows no sign of magnetic saturation up to 18$\,$T \cite{Fischer2018}. Thin films are more subject to crystallographic defects, resulting in pinning of the domain walls \cite{Xu2019} and thereby requiring larger magnetic fields to be manipulated. 

Domain walls also influence spin currents passing though NiO by reflection and absorption and they give rise to bound magnons. AFM domain walls can be engineered, induced and controlled via exchange bias \cite{Logan2012,Tveten2013}. 
These effects are more pronounced at low temperatures for which the activation energy to move the domain wall is larger than the thermal energy \cite{Michel1991}. The control over the magnetic moments in NiO can be increased using the exchange bias with YIG. At low temperatures, the exchange bias is large and comparable in size to the coercivity of YIG \cite{Li2016,Luan2018a}.

The SMR signal in bulk NiO increases with the etch step by a factor of two \cite{Hoogeboom2017} which could be caused by a change in the magnetic order. 
Velez et al. reported an etching step on a YIG sample leading to larger SMR signals \cite{Velez2016}, similar as with NiO. However, at low temperatures, the etching of YIG lead to a sign change of its SMR. This shows that the magnetic moment directions are aligned off the magnetic field direction which is, on average, larger than 45$\degree$. 

\begin{figure*}[t]
\includegraphics[width=18cm]{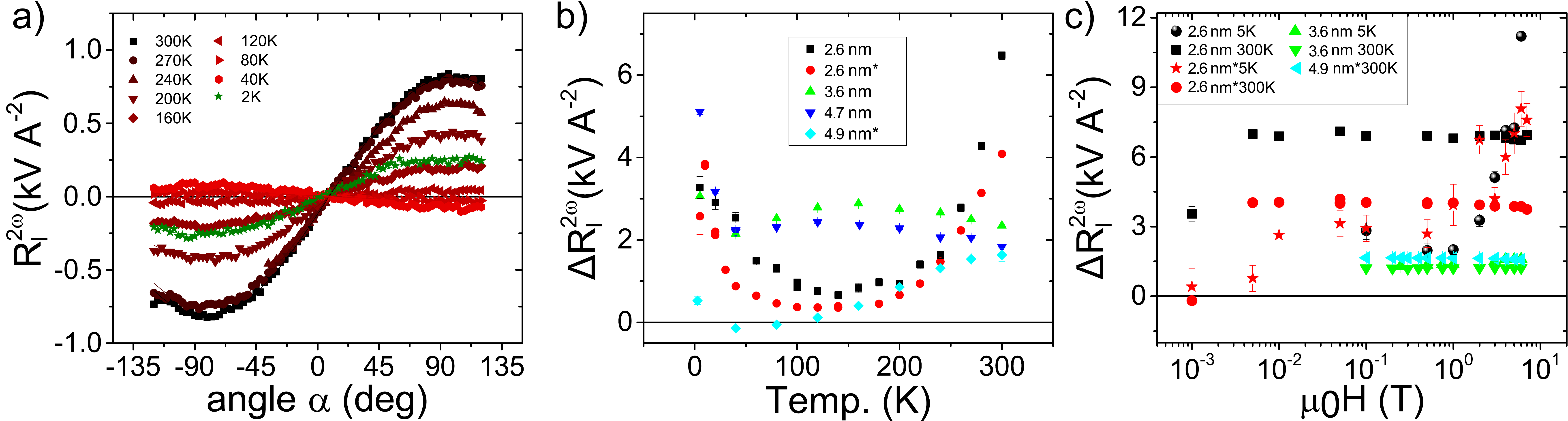}
\caption{Second harmonic local resistivity changes a) obtained for the Pt$|$NiO(4.9$\,$nm$^*$)$|$YIG sample at 2$\,$T and its fits as a function of the rotation angle for various temperatures. The SSE modulation of the resistance shows a phase shift of $180\degree$ at 40 and 80$\,$K equivalent to a sign change, while the 2$\,$K data shows a similar sign as the data obtained at temperatures at 120$\,$K and above. b) The amplitude of the sinusoidal fit $\Delta R^{2\omega}_l$ is depicted as a function of the temperature. c) The local SSE shows no magnetic field dependence at 300$\,$K. At 5$\,$K, however, an increasing magnetic field response for the 2.6$\,$nm samples is observed which is similar to the SMR results.}
	{\label{fig:local_SSE}}
\end{figure*}

The parameters for diffusion of spin currents through NiO are temperature dependent as well since this relies on thermally excited magnetic fluctuations \cite{Rezende2016a}. The magnetic fluctuations could also explain FM resonance measurements as a function of the NiO thickness \cite{Wang2014,Wang2015,Rezende2016a}. At low temperatures, an increase in the SSE signal from YIG has been observed, suggesting a strong correlation between magnon and phonon transport \cite{Iguchi2017}. In Pt$|$YIG$|$NiO$|$YIG$|$GGG structures a similar increase in SSE has been attributed to an increased NiO transmissivity and an increased contribution of the GGG substrate \cite{Chen2019}. When replacing the easy-plane AFM thin film with the uniaxial Cr$_2$O$_3$, normal spin currents can be blocked by the AFM film when the N\'eel vector lies perpendicular to the YIG magnetization \cite{Qiu2018}. 
Magnons crossing the interface of NiO$|$YIG might not experience a similar blocking mechanism as NiO is affected differently by the exchange bias and has many magnon modes extending to low frequencies \cite{Milano2010}, already populated at a few Kelvin. In addition to incoherent magnon transport, GHz magnons have been driven coherently through easy-plane AFM films, observed by means of element- and time-resolved x-ray pump-probe measurements \cite{Li2019,Dabrowski2020}. 

Spin transport theories consider the NiO as being an inactive, opaque layer. However, NiO is known to be responsible to create a SSE signal on its own. A SSE signal from a NiO layer of 200$\,$nm thickness on FM permalloy increases with increasing temperature starting from about 150$\,$K \cite{Ribeiro2018}. A magnon diffusion theory for the SSE in AFMs shows that the SSE is expected to go to zero at low temperatures in NiO \cite{Rezende2018}. However, a SSE signal was established on a $\mu$m length scale in bulk NiO with a nonlocal geometry at low temperatures \cite{Hoogeboom2020}. A magnetic field lifts the  degeneracy of low frequencies magnon modes with opposite spin, creating an imbalance in their population \cite{Cheng2014,Hoogeboom2020}.  
The observation of electrically injected spin currents through a thin film of $\alpha$-Fe$_2$O$_3$ whose relaxation length is governed by domain configurations \cite{Ross2020} showed a proof of principle of spin currents through single layer AFM thin films. 
The nonlocal geometry is employed for the NiO$|$YIG bilayer to investigate long-distance spin transport through thin NiO film as a function of the magnetic order affected by the temperature and the magnetic field.

\section{Results and Discussion}

Figure \ref{fig:SMR}a) shows the angular dependent SMR modulation of the Pt$|$NiO(4.9$\,$nm$^*$)$|$YIG device at 2$\,$T for various temperatures. A positive SMR is observed above 150$\,$K, while a 90° angular shift (negative SMR) can be identified for lower temperatures. The fitted amplitude for the devices with different NiO thicknesses as a function of temperature is shown in Fig. \ref{fig:SMR}b). The N\'eel temperature is an indication for the magnetic order and can be determined by a peak in the SMR or SSE temperature dependence \cite{Lin2016,Hou2017}. This is not observed within the temperature range except for the NiO(4.9$\,$nm$^*$) sample. A high N\'eel temperature is an indication of a high magnetic order. Bulk NiO has a N\'eel temperature of 525$\,$K and thinner films are expected to have a lower N\'eel temperature. However, the NiO(4.9$\,$nm$^*$) sample has a lower SMR sign change temperature than the 250$\,$K of the other samples. The etch step thus either lowers the order and thereby the NiO SMR contribution, or increases the transmissivity for spin currents towards and from YIG. The lack of such SMR peaks within the measured temperature range indicates that the N\'eel temperature is equal or higher than that of comparable thick NiO layered devices in literature \cite{Lin2016,Hou2017,Lin2017}.

\begin{figure*}[t]
\includegraphics[width=13cm]{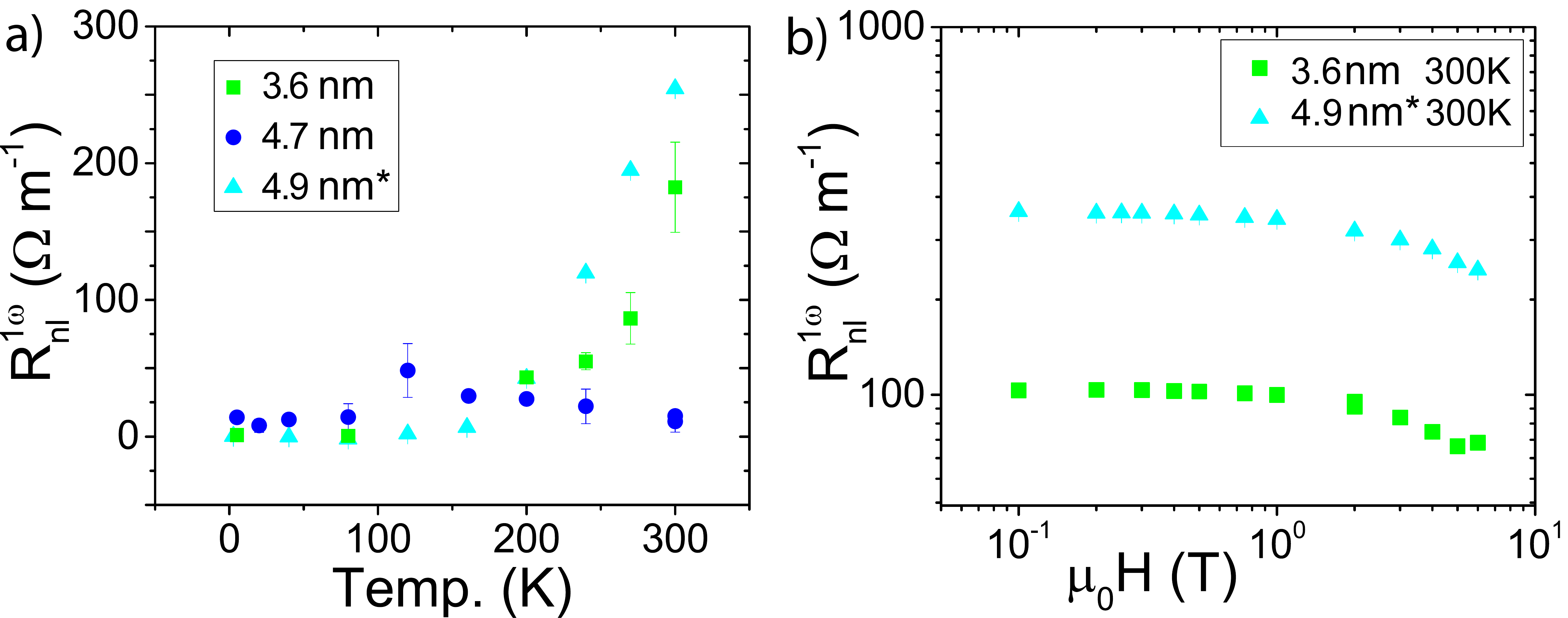}
\caption{Nonlocal electrically injected signal (1st harmonic) amplitude of different samples  with a distance $d$ between the Pt bars of about 750$\,$nm. 
The voltage is divided by the length of the Pt strips ($20\,\mu$m) as well as the injector current ($100\,\mu$A). a) The amplitude decreases by lowering the temperature showing a lowered transmissivity of NiO at low temperatures. b) As a function of the magnetic field strength, the room temperature signal slightly decreases, roughly following the same field dependence as for devices without the NiO interlayer.}
	{\label{fig:electrical_non-local}}
\end{figure*}

Figure \ref{fig:SMR}c) shows the SMR amplitude for different field strengths at either 5$\,$K and 300$\,$K. The amplitudes of all samples are fitted with an offset and a quadratic magnetic field dependence. The offset is a result of the exchange bias with YIG as all applied field strengths are larger than the sub mT coercivity of YIG, while the magnetic field strength required to manipulate a thin NiO film is 3 orders of magnitude higher \cite{Fischer2018}. The exchange bias of YIG therefore enables some control of the NiO magnetic moments even at the smallest applied magnetic field strengths. 

Both the Pt$|$NiO(7.5$\,$nm)$|$YIG and Pt$|$NiO(7.5$\,$nm$^{*}$)$|$YIG sample show a  negative quadratic SMR increase by further increasing the magnetic field, similar to the initial increase in SMR with field strength observed in Pt on bulk NiO \cite{Hoogeboom2017} and on thin NiO films \cite{Fischer2018}. With higher magnetic field strengths, the domain size increases and the Zeeman energy becomes larger than the magnetic anisotropy, allowing more control over the direction of the magnetic moments. The negative SMR increase with increasing magnetic field strength using AFMs is caused by the increased influence of the N\'{e}el vector in all magnetic domains \cite{Hoogeboom2017,Fischer2018}. Thinner NiO$|$YIG films, however, show a positive quadratic SMR increase with increasing magnetic field strength. Remarkably, a positive SMR sign is retrieved above 2$\,$T for the Pt$|$NiO(2.6$\,$nm$^*$)$|$YIG device at 5$\,$K. 

Since the positive SMR in Pt$|$NiO$|$YIG samples is attributed to the interaction of the accumulated spins at the Pt$|$NiO interface with YIG being dominant over the interaction with NiO \cite{Lin2017}. Thus, we can conclude that by increasing the magnetic field, not only the control over the magnetic moments increases, but also the spin current transmissivity becomes more efficient and the interaction with YIG becomes dominant. A higher magnetic order created with the magnetic field might increase the transmissivity. On top of this positive contribution, a positive quadratic increase could originate from the Hanle magnetoresistance, but this contribution is too small to be fully responsible for the sign change \cite{Dyakonov2007,Velez2016a,Shan2017a}.


Figure \ref{fig:local_SSE}a) shows the angular dependence of the local SSE signal in the Pt$|$NiO(4.9$\,$nm$^*$)$|$YIG device at 2$\,$T and various temperatures. Most curves follow a regular $\textbf{E}\propto \textbf{J}_s\times\textbf{M}$ dependence, except the curves at  80$\,$K and 40$\,$K. These curves show a 180$\degree$ angular shift in the signal, equivalent to a sign change. SSE sign changes have been reported in literature for ferrimagnetic gadolinium iron garnet, attributed to different contribution of the Gd and Fe sublattices at different temperatures \cite{Geprags2016}. 
This explanation does not hold for the NiO$|$YIG bilayers since the NiO moments are perpendicular to those of YIG \cite{Luan2018a,Dabrowski2020}.

The amplitude of the sinusoidal fits with a periodicity of 360$\degree$ is shown in Fig.  \ref{fig:local_SSE} b) for the different samples. Until about 150$\,$K, most of these amplitudes decrease with decreasing temperature. This decreasing transmissivity trend below the N\'eel temperature is observed in both SSE \cite{Lin2016} and SMR \cite{Lin2017,Hou2017} measurements. The NiO(3.6$\,$nm) and NiO(4.7$\,$nm) samples on the other hand show a more flat signal with temperature and have a smooth peak around 150$\,$K. Chen \textit{et al.} observed a similar peak for a 5$\,$nm interlayer, and attributed it to the combination of magnon population and relaxation in YIG, interface effects and an enhancements of spin currents near the blocking temperature around 30$\,$K \cite{Chen2019} and 50$\,$K \cite{Li2016}. No SSE signal is acquired within the noise for the NiO(7.0$\,$nm)$|$YIG device at room temperature, possibly because the layer is too thick for the exchange bias to affect the top part of the film. The NiO(4.9$\,$nm$^*$) sample, on the other hand, exhibits the same behavior as both 2.6$\,$nm samples. 

When further lowering the temperature, however, all samples show a recurring SSE signal. The size of these signals is comparable to the signal at room temperature. 
A higher thermal conductivity of the magnetic bilayer or a larger SSE coefficient could play a role in creating larger thermally created spin currents for the same given charge current in Pt. For bilayers including NiO, a SSE in NiO could be another source of enhanced spin current at low temperature. 

To study the sensitivity of these low temperature SSE signals to the magnetic order, field dependent measurements have been performed and compared to room temperature dependence, as shown in Fig. \ref{fig:local_SSE} c). 
At room temperature, no SSE increase is observed contrary to the SMR signal at these temperatures. The increased magnetic order related to the SMR increase thus seems to have little effect on the SSE induced spin current. 
At 5$\,$K there is, similar to the SMR signal, an increase with magnetic field. 

Several phenomena might affect these low temperature SSE signals; magnetic pinning by crystallographic defects, the enhanced exchange bias with YIG or NiO being a source of spin currents. To determine whether the spin transport at low temperatures has been improved or to establish that NiO is a spin current source, we have to compare these results with data of the nonlocal measurements.


\begin{figure*}[t]
\includegraphics[width=13.0cm]{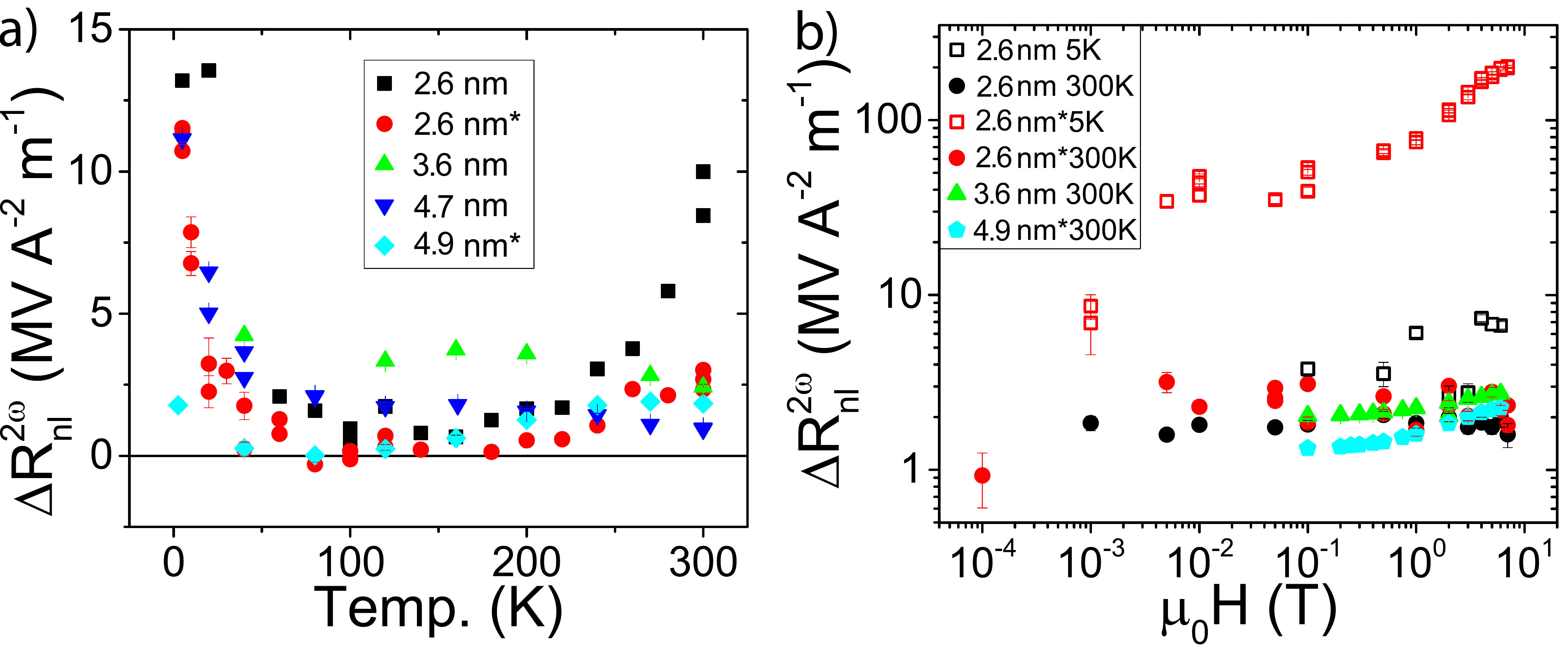}
\caption{Nonlocal thermally injected signal (2nd harmonic) amplitude for different samples. The voltage is divided by the length of the Pt strips ($20\,\mu$m) as well as the square of the injector current ($100\,\mu$A). a) When lowering the temperature below the N\'eel temperature, the signal decreases. Similar to the local SSE signal, the NiO(3.6$\,$nm) and NiO(4.7$\,$nm) samples show an additional peak around 150$\,$K. With further lowering of the temperature, all the samples show a recurring signal which is attributed to the thin NiO film. b) As a function of the field strength, the room temperature signal remains constant or slightly increases, while the 5$\,$K measurements show a larger increase with increasing field. In the etched sample this trend is amplified.}
{\label{fig:SSE_non-local}}
\end{figure*}

The room temperature nonlocal spin transport through the bilayer shows similar transport properties as without the NiO layer in terms of angular dependence, signal strength and field dependence \cite{Cornelissen2015,Cornelissen2016}. All signal minima and maxima correspond to the same direction of the magnetic field as observed in Pt$|$YIG systems with comparable Pt strip distances and are defined as positive. The electrical injection signals of the NiO(3.6$\,$nm) and NiO(4.9$\,$nm$^*$) samples in Fig. \ref{fig:electrical_non-local} a) show a signal which is a factor of 3 smaller than the signals obtained in YIG without the NiO film \cite{Cornelissen2015}. The NiO(4.7$\,$nm) sample shows considerable lower signals, possibly because the N\'eel temperature of the NiO layer is higher. When lowering the temperature of the NiO(3.6$\,$nm) and NiO(4.9$\,$nm$^*$) samples, a sharp signal decrease is obtained. 

Figure \ref{fig:electrical_non-local}~b) shows the nonlocal electrically injected signals as a function of magnetic field strength. A similar lowering is observed in YIG \cite{Cornelissen2016}. In the easy-plane anisotropy state, Hematite also shows a magnetic field dependence due to the rotation of the pseudo-spin, induced by the Dzyaloshinskii-Moriya interaction (DMI) which causes a net magnetization \cite{Wimmer2020}. Although DMI is not present in NiO, a rotation of the pseudo-spin might still occur with distance due to the easy-plane anisotropy \cite{Kamra2020}. The exchange bias might even act as a DMI as it results in a net magnetization. This would be fairly small since the exchange field is less that 1$\,$mT \cite{Luan2018a}. The small effect of the magnetic field shows that diffusion of spin currents through the NiO also does not strongly depend on magnetic ordering of NiO at room temperature. 

The nonlocal SSE also shows a decrease in signal with decreasing temperature (Fig. \ref{fig:SSE_non-local}~a)), although not as pronounced as the electrically injected signal. This can be explained by the different path of these magnons; the electrically injected magnons at the Pt$|$NiO interface passes the NiO twice while the largest part of the heat induced spin current will be created within the thicker YIG layer. Because the gradient is radially distributed, most of the spin current is generated in the vicinity of the injector, after which the NiO only needs to be passed once in order to be detected.

At low temperatures, the nonlocal SSE shows similar recurring signals as measured locally. The size of the signals is even larger at 5$\,$K than at 300$\,$K. Figure \ref{fig:SSE_non-local}~b) shows the nonlocal thermally generated signals as a function of the magnetic field strengths at 5$\,$K and 300$\,$K. At room temperature there is little effect of the magnetic field strength, but the small effect shows increasing SSE signals as a function of the magnetic field strength. At low temperatures, however, a considerable increase with magnetic field strength is observed.

Generally, there is a decrease in spin transport through NiO by lowering the temperature from room temperature to about 150$\,$K, which is not observed in samples without a NiO interlayer \cite{Cornelissen2016a}. This can be explained by the lowered transport governed by diffusion in this temperature region \cite{Rezende2018} due to a lower amount of magnetic fluctuations. At low temperatures, however, the nonlocal SSE signal deviates from the present understanding of lowered transmissivity at low temperatures with an inactive NiO layer. First of all, there are some observations of negative SSE values around 40$\,$K to 80$\,$K locally and nonlocally. Even though the transmissivity for spin transfer torque induced spin currents decreases at low temperatures, the SSE results of both local and nonlocal geometries show an increase in signal at low temperatures which increase with increasing field. 

Unlike the room-temperature SSE and the nonlocal electrically injected magnons, SMR signal strength seems to increase with increasing magnetic field strength. The increasing magnetic order with larger magnetic field strength might influence the transmissivity of the NiO layer, increasing the spin transfer torque exerted on the YIG. The effect of domain wall pinning by defects is stronger at low temperatures and larger fields might be required for control over the NiO magnetic moments. Contrary to the local SSE signals, the room temperature nonlocal SSE signals do increase with increasing field. Moreover, the field dependence at 5$\,$K is strong enough to make the SSE signal exceed the 300$\,$K signals at large fields for both the local and nonlocal geometries. 

The effect of magnetic order on the spin transport is significant and influenced by etching. In addition to the SMR sign change with field at 5$\,$K, the nonlocal SSE of the NiO(2.6$\,$nm$^*$)$|$YIG device shows an increase of more than one order of magnitude in signal strength while the signal of the non-etched sample remains more constant with field. The etch step affects the interface by cleaning it from surface adatoms and by affecting the magnetic order. The etching does influence the NiO(4.9$\,$nm$^*$)$|$YIG sample, showing considerably lower SSE signals than the NiO(4.7$\,$nm)$|$YIG sample. This NiO(4.9$\,$nm$^*$)$|$YIG sample is etched only below the Pt strips. Since the NiO(2.6$\,$nm$^*$)$|$YIG sample was afterwards exposed to ambient conditions for lithography purposes the cleaning seems less relevant. The local SSE shows little effect of the etch, indicating that the etch is most relevant for a spin current flowing through the NiO layer at low temperature.


Similar to these NiO$|$YIG$|$GGG samples, a low temperature upturn in nonlocal SSE signal has been observed for the paramagnet substrate GGG itself \cite{Oyanagi2019}. The signal size at a distance $d\,=\,$500$\,$nm is around 5$\,$k V A$^{-2}$ m$^{-1}$. This is 3 orders of magnitude smaller than the signals observed in the NiO$|$YIG$|$GGG geometry. The GGG thus plays a negligible role in the SSE upturn at low temperatures. For YIG films without the NiO interlayer, peaks in the SSE have been observed at low temperatures \cite{Cornelissen2016a}. An increase in the SSE coefficient of YIG is held responsible for the increase in the signal, but the mechanism is not well understood. 

The SSE signal can be the result of different sources for the spin current. In YIG, a decrease of the SSE signal observed with increasing magnetic field strength \cite{Cornelissen2016} while an increase is observed with the NiO interlayer. Furthermore, the NiO in expected to become less transmissive for spin currents originating from YIG at low temperatures. Therefore, an active role of NiO as a source of spin currents could be responsible for the upturn of the SSE signals at low temperatures. The low temperature SSE signals observed in these NiO$|$YIG samples indeed show similarities to the SSE observed in bulk NiO. The increase at low temperatures resembles the increase of SSE in bulk NiO \cite{Hoogeboom2020}. Moreover, the temperature for which the signal arises and the dependence on the magnetic field strength is similar to bulk NiO, shown to be originating from the imbalance in the population of the magnon modes with opposite spin \cite{Hoogeboom2020}.

Initial spin pumping \cite{Wang2014} and SSE \cite{Lin2016} signals at room temperature increase have been observed by inserting up to a few nm of NiO film between the Pt and YIG. 
This has been attributed to an enhanced spin conductivity \cite{Li2019,Dabrowski2020}. The further exponential decay of SSE signal strength with increasing NiO thickness might be influenced by the increase in N\'eel temperature of thicker films, shifting the peak in transmissivity towards higher temperatures. At constant temperature, the spin current transmissivity then decreases for thicker NiO films. Nonetheless, the peak height of the SSE signal also decreases with increasing NiO thickness \cite{Lin2016,Prakash2016}.

However, the low-temperature SSE signal is not related to the thickness of the NiO layer, indicating to be unrelated to the transmissivity of spin currents from YIG. 
The passive, diffusive role for spin currents in NiO is shown to be minimally depending on the magnetic field strength in case of the electrically injected magnons. Although the transmissivity seems unrelated to the source of the spin currents, the SSE does increase significantly at low temperatures. Therefore, we attribute the low temperature SSE signal to the NiO itself.

\section{Conclusion}
Spin currents have been injected by electronic and thermal means into NiO$|$YIG samples with various thicknesses at a wide range of temperatures and magnetic field strengths. The spin current transmissivity of NiO peaks at the N\'eel temperature and is reduced by lowering the temperature. 
At low temperatures, however, there is a recurring thermally generated spin current which has been detected both locally and nonlocally. An increase in thermal conductivity could affect thermally generated spin current in this temperature regime. However, the low temperature SSE signals are not affected by the NiO thickness and therefore seem unrelated to possible changes in the transmissivity of the NiO layer. On the other hand, the low temperature SSE signals resemble those observed in bulk NiO: increasing signal strengths with increasing magnetic field strength and decreasing temperature. This indicates that, in addition to the passive, diffusive role, the NiO plays an active part in the SSE signals by generating thermal spin currents itself at low temperatures.

\section{Aknowledgements}

We acknowledge J. G. Holstein, H. Adema, T. J. Schouten and H. H. de Vries for their technical assistance. In addition, we thank Martin Gottschalk and Karsten Rott for support and discussion regarding the TEM experiments as well as Andreas Hütten and Günter Reiss for making available the laboratory equipment for sample characterization. This work is part of the research program Magnon Spintronics (MSP) No. 159 financed by the Nederlandse Organisatie voor Wetenschappelijk Onderzoek (NWO). Further support by the DFG Priority Programme 1538 “Spin-Caloric Transport” (KU 3271/1-1) and the Spinoza Prize awarded in 2016 to B. J. van Wees by NWO is gratefully acknowledged.

\bibliography{references}

\end{document}